\documentclass[journal,twoside,web]{ieeecolor}
\makeatletter
\let\NAT@parse\undefined
\makeatother
\usepackage{tmi}
\usepackage{cite}
\usepackage{amsmath,amssymb,amsfonts}
\usepackage{algpseudocode}
\usepackage[ruled,linesnumbered]{algorithm2e}
\usepackage{graphicx}
\usepackage{textcomp}
\usepackage{diagbox}
\usepackage{tabularx}
\usepackage{multirow}
\usepackage[colorlinks,linkcolor=blue,anchorcolor=blue,citecolor=green]{hyperref}

\def\BibTeX{{\rm B\kern-.05em{\sc i\kern-.025em b}\kern-.08em
    T\kern-.1667em\lower.7ex\hbox{E}\kern-.125emX}}
\newcommand{\tabincell}[2]{\begin{tabular}{@{}#1@{}}#2\end{tabular}} 
\begin{document}
\title{GASNet: Weakly-supervised Framework for COVID-19 Lesion Segmentation}
\author{Zhanwei Xu, Yukun Cao, Cheng Jin, Guozhu Shao, Xiaoqing Liu, \\Jie Zhou, \IEEEmembership{Senior Member, IEEE}, Heshui Shi, Jianjiang Feng, \IEEEmembership{Member, IEEE}
\thanks{This work was supported in part by the National Natural Science Foundation of
 China under Grant 82071921 and Zhejiang University special scientific research fund for
  COVID-19 prevention and control.}
\thanks{Zhanwei Xu, Cheng Jin, Jianjiang Feng, and Jie Zhou are with Department of Automation, 
Beijing National Research Center for Information Science and Technology, 
Tsinghua University, Beijing 100084, China 
(e-mail: xzw14@tsinghua.org.cn, orangeking2020@gmail.com, jfeng@tsinghua.edu.cn, jzhou@tsinghua.edu.cn).}
\thanks{Yukun Cao, Guozhu Shao, Xiaoqing Liu and Heshui Shi are with Department of Radiology, Union Hospital, 
Tongji Medical College, Huazhong University of Science and Technology, 
and Hubei Province Key Laboratory of Molecular Imaging, Wuhan 430022, China 
(e-mail: caoyukun@foxmail.com, 1946487947@qq.com, lxq\_xh@163.com, heshuishi@hust.edu.cn).}
\thanks{Z. Xu and Y. Cao contributed equally to this work. 
Corresponding authors. Jianjiang Feng (jfeng@tsinghua.edu.cn) and Heshui Shi (heshuishi@hust.edu.cn).}
}

\maketitle

\begin{abstract}
Segmentation of infected areas in chest CT volumes is of great significance for further diagnosis and treatment of COVID-19 patients.
Due to the complex shapes and varied appearances of lesions, 
a large number of voxel-level labeled samples are generally required to train a lesion segmentation network,
which is a main bottleneck for developing deep learning based medical image segmentation algorithms.
In this paper, we propose a weakly-supervised lesion segmentation framework by embedding
 the Generative Adversarial training process into the Segmentation Network, which is called GASNet.
GASNet is optimized to segment the lesion areas of a COVID-19 CT by the segmenter, and to replace the \textit{`abnormal'} appearance with a generated \textit{`normal'} appearance 
by the generator, so that the `restored' CT volumes are indistinguishable from healthy CT volumes by the discriminator.
GASNet is supervised by chest CT volumes of many healthy and COVID-19 subjects without voxel-level annotations.
Experiments on three public databases show that when using as few as one voxel-level labeled sample,
the performance of GASNet is comparable to fully-supervised segmentation algorithms trained on dozens of voxel-level labeled samples.

\end{abstract}

\begin{IEEEkeywords}
   Convolution Neural Network, COVID-19, GAN, Weakly-supervised Segmentation
\end{IEEEkeywords}

\section{Introduction}
\begin{figure}[ht]
    \centering
    \includegraphics[scale = 0.9]{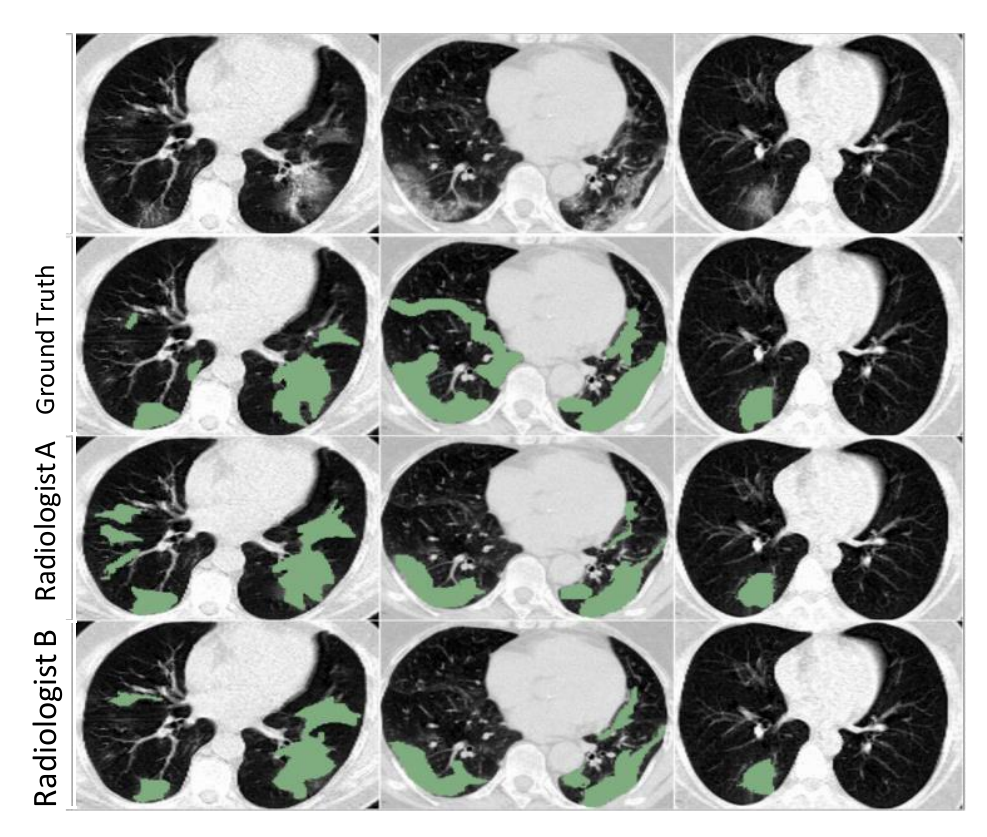}
    \caption{Typical CT scans of three COVID-19 patients from the public Dataset-A~\cite{COVID-19-SegBenchmark}. 
    From the second row to the fourth row are the different annotations provided along with the dataset and by two other radiologists from Wuhan Union Hospital.
    The difference between different annotations is obvious.
    }
    \label{fig:typical}
\end{figure}

\IEEEPARstart{T}{he} epidemic of coronavirus disease 2019 (COVID-19) is raging around the world. 
Chest CT can detect small lesion areas due to its high spatial resolution and therefore is an effective imaging tool for monitoring the disease~\cite{fang2020sensitivity}\cite{kanne2020chest}.
Automatic segmentation of lesion areas of a COVID-19 CT can facilitate medical experts in diagnosing by focusing on the Region of Interest (RoI) instead of the whole volume.
Besides, statistics related to the lesion area
are part of the criteria for determining the severity~\cite{chung2020ct}\cite{shi2020radiological}.
However, this task is challenging as 
the lesion areas are extremely varied.
Three typical COVID-19 CT scans from a public dataset~\cite{COVID-19-SegBenchmark} are shown in Fig.~\ref{fig:typical}. 
It can be seen that the lesions range from small to large, 
and the appearance may be of ground glass opacity, consolidation, or mixed type.
Due to blurry and indistinguishable boundaries between infected and healthy areas, voxel-level labeling of lesions is not only time-consuming,
but also tends to contain inconsistency between different annotators.
Fig.~\ref{fig:typical} also shows the infectious annotation as ground truth (GT) along with the dataset and the manual segmentation results by two other radiologists. 
Since the boundary of the infected area is very fuzzy, 
even the segmentation results given by two experienced radiologists have obvious inconsistencies with GT.

Deep learning shown encouraging performance for lesion segmentation of COVID-19 CT,
but only when a sufficient amount of labeled data such as thousands of slices is available~\cite{zhang2020clinically}\cite{wu2020jcs}\cite{shan2020lung}\cite{qiu2020miniseg}.
It takes more than 200 minutes on average to annotate the lesion area of one COVID-19 CT volume~\cite{shan2020lung}. 
The high cost of collecting expert annotations is a big obstacle to the development of medical image segmentation algorithms for new diseases like COVID-19.

Data augmentation~\cite{cciccek20163d}\cite{zhang2019unseen} and image synthesis~\cite{shin2018medical}\cite{jin2018ct} may alleviate the lack of pixel/voxel-level annotations to a varying degree.
Self-learning or active learning~\cite{bai2017semi}\cite{nie2018asdnet}\cite{zhang2018self} updates the segmentation model by iteratively providing pseudo label to 
the unlabeled data,
and hopes to gradually improve the precision.
Other methods try to make up for the lack of pixel/voxel-level supervision information by using image/volume-level labels, such as
Class Activation Maps (CAMs)~\cite{selvaraju2017grad}, Generative Adversarial Network (GAN)~\cite{radford2015unsupervised},
and Multiple Instance Learning (MIL)~\cite{zhou2009multi}.
However, these methods have more or less the following problems:
(1) a certain number of samples with pixel/voxel-level annotations are still necessary;
(2) using pseudo-label data 
may introduce noise;
and (3) the mapping from volume-level annotation information to voxel-level segmentation is usually not accurate enough.
Detailed comment on related medical image segmentation methods is provided in section~\ref{sec:related}.

\begin{figure*}[]
    \centering
    \includegraphics[scale = 0.23]{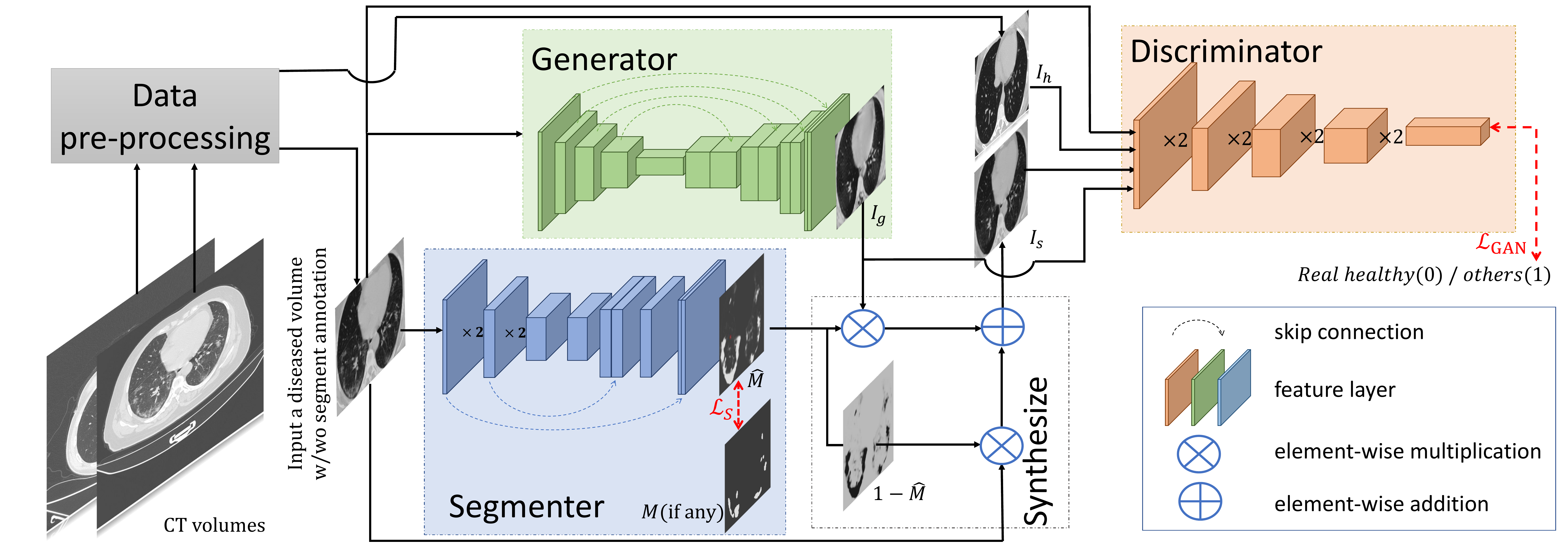}
    \caption{Three modules with optimizable parameters compose the framework of GASNet, the segmenter (S), the generator (G), and the discriminator (D). 
      Only the S part is needed during the test.
      The input of the pipeline is presented in 2D for clarity. 
      In fact, the input to each module of GASNet is the entire 3D volume.}
    \label{fig_pipeline}
\end{figure*}

Our idea is to `restore' the CT volume of a COVID-19 patients to the status when he/she is healthy by combining a segmentation network and a generative network.
Restoration performance is supervised by a discriminator that is trained using CT scans of many healthy people and COVID-19 patients (without voxel-level labeling of lesion areas).
This scheme is feasible since a large number of volume-level labels, indicating
whether a CT volume is COVID-19 positive or not, are directly available from diagnosis results in COVID-19 designated hospitals 
and more reliable~\cite{jin2020development} than voxel-level annotations obtained manually.
The proposed framework is designed to mine the potential knowledge contained in many COVID-19 positive and negative CT volumes
by embedding Generative Adversarial training in a standard Segmentation Network, referred to as GASNet,
and hence its demand for voxel-level annotations is very small.
Fig.~\ref{fig_pipeline} shows the pipeline of GASNet. Both the generator and the segmenter take a COVID-19 CT volume as input, and
the two outputs together with the original CT volume are fused to form a synthetic healthy volume. 
Both real and synthetic healthy volumes are fed to the discriminator.
In the training process, the goal of the discriminator is to distinguish between the synthetic healthy volume and the real healthy volume,
while the goal of the generator and the segmenter is to deceive the discriminator.
Such an adversarial training strategy will push the segmenter to segment the lesion areas of a COVID-19 CT as precisely as possible.
We also propose a simple but effective strategy of synthesizing COVID-19 CT volumes with voxel-level pseudo-labels during the adversarial training process, 
which further improves the segmentation performance of GASNet.
A detailed description of the algorithm will be given in section~\ref{sec:method}.

Compared with other weakly-supervised methods, a major advantage of GASNet lies in utilizing volume-level labels in an adversarial learning way, 
alleviating the burden of voxel-level annotation and maintaining a good segmentation performance at the same time. 
When using only one voxel-level labeled sample in training, GASNet obtains a 70\% Dice score on a public COVID-19 lesion segmentation dataset~\cite{COVID-19-SegBenchmark}, 
comparable to representative fully-supervised algorithms (U-Net~\cite{ronneberger2015u}, V-Net~\cite{milletari2016v}, and UNet$^{++}$~\cite{zhou2018unet++}) requiring a large number of voxel-level annotated samples. 
Code of GASNet is available at \url{https://github.com/xzwthu/GASNet}.
Details of the experiments are in section~\ref{sec:experiment} and section~\ref{sec:ablation}. 

\section{Related Work}
\label{sec:related}
In this section, we first introduce existing public COVID-19 CT databases containing lesion annotation, 
then review current COVID-19 lesion segmentation methods both in fully-supervised and weakly-supervised way, 
and finally describe recent weakly-supervised methods and GAN methods in general medical image segmentation.
\subsection{Public COVID-19 CT segmentation datasets}
Performance evaluation using public datasets is very important for comparing different image segmentation algorithms.
Being an emerging research direction, most COVID-19 studies are conducted independently~\cite{zhang2020clinically}\cite{wu2020jcs}\cite{shan2020lung}, using non-public data.
Very recently, a few public databases are available~\cite{COVID-19-SegBenchmark}\cite{MedSeg}\cite{morozov2020mosmeddata}.
Giving a brief description of these databases is necessary before we discuss performance of different algorithms.
We summarize the current public COVID-19 CT segmentation datasets in Table~\ref{tab:dataset}. Detailed descriptions are in section~\ref{sec:experiment}.
\begin{table}[htb]
    \caption{A summary of public COVID-19 CT datasets with lesion annotations.}
    \centering
    \begin{tabular}{ccc}
    \hline
    \tabincell{c}{Dataset}  & \tabincell{c}{ Voxel-level\\annotations} &\tabincell{c}{ Brief} \\
    \hline
    \tabincell{c}{A~\cite{COVID-19-SegBenchmark}}  & \tabincell{c}{20} &\tabincell{c}{CT volumes from two sources, containing left\\lung, right lung, and COVID-19 lesion masks.}\\
    \hline
    \tabincell{c}{B~\cite{MedSeg}}  & \tabincell{c}{9} & \tabincell{c}{CT volumes from Radiopaedia~\cite{radiopaedia}, containing \\lung masks and COVID-19 lesion masks.}\\
    \hline
    \tabincell{c}{C~\cite{morozov2020mosmeddata}}  & \tabincell{c}{50} & \tabincell{c}{CT volumes of mild patients, \\containing COVID-19 lesion masks.}\\
    \hline    
    \end{tabular}
    \label{tab:dataset}
\end{table}
\subsection{Fully-supervised COVID-19 lesion segmentation}

Most of the COVID-19 lesion segmentation methods~\cite{zhang2020clinically}\cite{wu2020jcs}\cite{shan2020lung} are based on U-Net~\cite{ronneberger2015u} structure or its modifications,
containing an encoding path and a decoding path, which are connected by skip connection at the corresponding resolution.
Zhang et al.~\cite{zhang2020clinically} adopt a two-stage segmentation framework for segmenting lung lesions into five classes.
They train on a total of 4,695 CT slice images with voxel-level annotations and obtain an mDICE score of 58.7\%. 
Wu et al.~\cite{wu2020jcs} jointly train a segmentation network and a classification network,
using over 144K slices including 3,855 voxel-level labeled CT scan slices from 200 COVID-19 patients.
They obtain a 78.3\% Dice score on their dataset.
Shan et al.~\cite{shan2020lung} use a 3D VB-Net as the backbone and employ a Human-In-The-Loop (HITL) strategy to train the network on 400 CT volumes. 
The HITL strategy reduces the annotation time and improves the accuracy. 
The net time spent on labeling data is still more than 176 hours, and they report a 91\% Dice score on their own dataset.
However, ~\cite{wu2020jcs}\cite{shan2020lung} have not published their codes, neither did they report their performance on public datasets.
We reproduce the network structures of these works and test their performance on Dataset-A~\cite{COVID-19-SegBenchmark} 
following a 5-fold cross-validation strategy.
The performance is slightly improved compared with a normal U-Net network, with Dice scores of 64\% and 63\%, 
while the performance of U-Net is 62\%. For more details, please refer to section~\ref{sec:ablation}. 

Besides U-Net, other deep models have also been used for COVID-19 lesion segmentation.
Fan et al.~\cite{fan2020inf} propose modules named Parallel Partial Decoder and Reverse Attention Module to improve lesion segmentation performance. 
They also conduct a test with semi-supervised strategy,
 collecting an unlabeled dataset and giving pseudo values iteratively, 
and gain a Dice score of 59.7\% on Dataset-B~\cite{MedSeg}. 
Qiu et al.~\cite{qiu2020miniseg} propose a lightweight 2D model pre-trained on ImageNet dataset~\cite{krizhevsky2012imagenet} and obtain performance comparable to heavy models like the fully convolutional network (FCN)~\cite{long2015fully} structure (77\% VS 75\%)
on a dataset consisting of 110 axial CT slices from $\sim$60 patients with COVID-19~\cite{SIRM}.  

\subsection{Weakly-supervised COVID-19 lesion segmentation}
\label{subsec:weaklyCOVID}
The latest research begins to explore lesion segmentation of COVID-19 volumes in weakly-supervised scenarios. 
Laradji et al.~\cite{Laradji2020} propose to train a neural network with active learning on a point-level annotation scenario.
Yao et al.~\cite{Yao2020} 
design a set of simple operations to synthesize lesion-like appearances, 
generate paired training datasets by superimposing synthesized lesions on the lung regions of healthy images,
and train a model to predict the healthy lung part of the input. 
A set of specially designed methods combining threshold selection, morphological processing, and region growth are used to determine the lesion segmentation during the test.
Zhang et al.~\cite{Zhang2020a} also use the GAN network as we do, but the purpose of GAN in their method is to perform data augmentation based on existing voxel-level labeled samples, 
so as to generate more paired samples with pseudo labels. 
Two segmentation networks, ENet~\cite{paszke2016enet} and U-Net, are trained to verify the effectiveness of the proposed data augmentation.

\begin{table}[htb]
    \caption{Published studies on weakly-supervised COVID-19 lesion segmentation.}
    \centering
    \begin{tabular}{c|c|c|c|c}
    \hline
    \tabincell{c}{Method} & Dataset &\tabincell{c}{Training volumes \\number with \\pixel/voxel-level \\annotation} & \tabincell{c}{Testing\\volumes\\number} &\tabincell{c}{Dice\\score\\(\%)} \\
    \hline
    \tabincell{c}{ActiveLearning~\cite{Laradji2020}}
    & A~\cite{COVID-19-SegBenchmark} & \tabincell{c}{16} & \tabincell{c}{4    }  & \tabincell{c}{52.4}  \\
    \hline
    \tabincell{c}{LabelFree~\cite{Yao2020}    }   
     & B~\cite{MedSeg} & \tabincell{c}{0}  & \tabincell{c}{8}      & \tabincell{c}{$59.4$}  \\ 
    \hline
    \tabincell{c}{CoSinGAN~\cite{Zhang2020a}} 
    & A~\cite{COVID-19-SegBenchmark}& \tabincell{c}{2} & \tabincell{c}{18 }              & \tabincell{c}{$57.8$} \\
    \hline
    \tabincell{c}{GASNet(ours)} 
     & A~\cite{COVID-19-SegBenchmark}& \tabincell{c}{1} & \tabincell{c}{19}    & \tabincell{c}{$\bf{70.3}$} \\
    \hline    
    \end{tabular}
    \label{tab:comparison}
\end{table}
Different from the above methods, we focus on designing a weakly-supervised segmentation framework under volume-level label supervision. 
Our framework simultaneously trains the 
GAN and the segmentation network and dynamically extracts the volume-level annotation information through adversarial learning, thus minimizing the requirement for voxel-level annotations.  
The comparison of GASNet with the above methods on the division of dataset, the number of annotations, and performance of segmentation are given in Table~\ref{tab:comparison}, 
and a more detailed comparison will be given in section~\ref{sec:experiment}.

\subsection{Weakly-supervised medical image segmentation}
\label{subsec:weakly}
Various methods of using weak annotations have been proposed in medical image segmentation area. 
Several works are devoted to the use of extra but sparse annotations, including scribbles~\cite{Valvano},
 points~\cite{wang2019weakly}\cite{Qu2019},
 and bounding boxes~\cite{Dolz2020}\cite{Rajchl2016DeepCut}. Scribbles and points require labeling at least one scribble or point for each RoI and
the labeled areas will be used to calculate the segmentation loss directly. As for the unlabeled part, Wang et al.~\cite{wang2019weakly} propose generating initial 
segments via a random walker algorithm~\cite{Grady2006Random}, and then train a fully-supervised segmentation network. Qu et al.~\cite{Qu2019} propose a similar pipeline
using a different method for label generation, combining K-means clustering results and Voronoi partition diagram. Instead of generating a pseudo label for the unlabeled areas, 
Valvano et al.~\cite{Valvano} directly predict the segmentation results by adding shape constraints through multi-GAN to make the segmentation results look realistic at multi-scales.
 Bounding boxes provide a more well-refined position constraint for segmentation but are more time-consuming for annotation~\cite{Dolz2020}\cite{Rajchl2016DeepCut}.

The major limitation of the aforementioned approaches is
relying on additional dataset annotations, which can be time-consuming and is prone to errors 
(for example, not all voxels in the bounding box should be positive; scribbles and points annotation can miss challenging labeled samples), and the errors can be propagated to the models during training. 
Methods using GAN, such as~\cite{Valvano} even need unpaired real segmentation masks, which are voxel-level labeled, as the real samples for the discriminator.

Weakly-supervised learning under volume-level label supervision earns increasing interest in medical image segmentation because it adds no annotation burden.
Xu et al.~\cite{Xu} enrich the volume-level labels to instance-level labels by multiple instance learning (MIL) and segment histopathology images
using only volume-level labels. However, MIL shows unsatisfactory performance on lesion segmentation of COVID-19 as shown in section~\ref{sec:experiment}. 
Feng et al.~\cite{Feng2017} propose a method especially for pulmonary nodules segmentation that learns discriminative regions from the activation maps of convolution units (CAM) in an image 
classification model. Ouyang et al.~\cite{Ouyang2019} employ the attention masks derived from a volume-level classification model as the voxel-level masks for 
weakly-annotated data. Because the attention masks are rough and inaccurate, hundreds of voxel-level annotations are still necessary for accurate lesion segmentation
like pneumothorax segmentation in chest X-ray~\cite{Ouyang2019}.

\subsection{GAN for medical image segmentation}
GAN is increasingly adopted as an assistance to medical image segmentation task.
The mainstream directions of GAN based methods include:
(1) synthesizing more available training sample pairs~\cite{shin2018medical}\cite{jin2018ct}\cite{mahapatra2018efficient},
 where GAN is a tool for data augmentation, 
and the training of segmentation network has no feedback on the quality of synthetic data.
(2) Adapting domain to leverage external labeled datasets~\cite{zhang2018task}\cite{chartsias2017adversarial}\cite{Valvano}.
The external dataset is required to contain enough pixel/voxel-level labeled training samples.
And (3) considering the segmentation network as a generator
and designing the discriminator as a structure of FCN~\cite{long2015fully} to obtain a confidence map of segmentation prediction, 
and thus helping the optimization of the segmentation network based on it~\cite{nie2018asdnet}\cite{mirikharaji2019learning}\cite{nie2020adversarial}.
Such methods do not use volume-level annotation, and their requirement for voxel-level labeled samples is considerable.

\section{Proposed Method}
\label{sec:method}
In this section, we first illustrate the pipeline of GASNet.
Then, we describe the auxiliary constraint terms in the form of loss functions used to make the training more stable and GASNet perform better.
We will also detail a simple but effective method of generating COVID-19 positive CT volumes with voxel-level pseudo-label 
to improve the segmentation performance of GASNet. 
Finally, we provide the implementation details,
including the specific structure,
data preprocessing, and the training hyperparameters.

\subsection{GASNet}
GASNet consists of three modules:
the generator (G), the discriminator (D), and the segmenter (S).
The data input to GASNet includes a small amount of voxel-level labeled data $I_l$,
and a large amount of volume-level labeled data $I_d$ and $I_h$, where $I_d$ is the diseased volume data and $I_h$ is the healthy volume data.
Our method is based on a simple fact:
the appearance of a lesion area contains the most obvious feature to distinguish COVID-19 CT from healthy CT.
We train a segmenter that can provide segmentation masks and utilize a generator to replace the predicted lesion area with a generated one 
whose appearance is close to the uninfected area while maintaining the uninfected area. 
If the synthetic healthy volumes successfully deceive the binary classifier, which is the discriminator in GASNet, 
we can obtain an accurate enough segmentation result.
The synthetic volume is fomulated by:
\begin{equation}
    I_s= \phi(S,G,I_d;\theta_S,\theta_G) = \hat{M}\times I_g+ (1-\hat{M})\times I_d, \label{eq1}
\end{equation}
where $\hat{M}=S (I_d;\theta_S)$ is the probabilistic segmentation mask predicted by S, $I_g = G (I_d; \theta_G)$ is the generated volume, 
and $\theta_S$, $\theta_G$ are the learnable parameters of S and G respectively.

To fully deceive the discriminator,
the segmenter needs to segment all infected areas and the generator needs to generate confusing volumes at the predicted lesion area of the segmentation.
In contrast, the discriminator will try to distinguish the synthetic volume from the real healthy one. We label the synthetic volume as 1 and
the real healthy volume as 0, and train the GASNet in an adversarial way via the following minimax game: 
\begin{equation}
    \mathop{\text{min}}_{\theta_{G},\theta_S} \mathop{\text{max}}_{\theta_D} \mathcal{L}_{GAN}(G,D,S) 
\end{equation}
where the objective function $\mathcal{L}_{GAN}$\footnote{
    We denote $\mathcal{L}_{GAN}\triangleq \mathcal{L}_{GAN}(G,D,S)$, $\mathop{\mathbb{E}_{\bf{I_h}}\triangleq \mathbb{E}_{I_h\sim p_{data}(I_h)}}$ and 
    $\mathop{\mathbb{E}_{\bf{I_d}}\triangleq \mathbb{E}_{I_d\sim p_{data}(I_d)}}$ for simplicity. 
} is given by
\begin{eqnarray}
    \mathcal{L}_{GAN}
    = \mathbb{E}_{\bf{I_h}}[\text{log} (1-D (I_h;\theta_D))] 
    +\mathbb{E}_{\bf{I_d}}[\text{log} (D (I_s;\theta_D))]  \nonumber 
\end{eqnarray}
where $D (I;\theta_D)$ is the prediction of the D, $I_s$ is fomulated by Eq~\ref{eq1}, and $\theta_D$ represents the learnable parameters of D.

As the formation of the synthetic volume is related to the prediction mask and the generated volume, gradient of $\mathcal{L}_{GAN}$ can feed back to 
both the S and G. Also, we add a basic segmentation loss  measuring the difference between the output of S and the GT of a small number of voxel-level labeled samples:
    $\mathcal{L}_S\footnote{
        We denote $\mathcal{L}_{S}\triangleq \mathcal{L}_{S}(S)$ and CEL(·,·) $\triangleq$ CrossEntropyLoss(·,·) for simplicity.
    }=\text{CEL} (\hat{M}_l,M_l)$,
where $\hat{M}_l=S (I_l;\theta_S)$, $M_l$ is the ground truth of the voxel-level labeled data $I_l$.

\subsection{Auxiliary constraints in the form of loss functions}
\label{sec:prior constrain}
Logically and theoretically,
provided that  we carefully train G, D, and S,
the synthesized volume will be nearly close to the healthy volume.
However, frameworks with GAN are generally difficult to train~\cite{nie2020adversarial}\cite{salimans2016improved}\cite{yu2017unsupervised}\cite{brock2018large}.
The quality of the generator and the discriminator is the crux for 
our ultimate goal of segmenting infected areas accurately.
Several auxiliary constraints are added to the loss functions to make the adversarial training more stable, leading to better performance.


First, the naive GASNet contains defects of the bias of the input. The segmenter is fed with only diseased volumes in the original GASNet. This brings sample bias
 and leads to false-positive predictions on healthy samples during testing.
For healthy CT volumes, we expect the predicted segmentation maps are all zero, and the output of the generator is a reconstructed volume of the original input. 
Therefore, the healthy volumes are also inputted into the segmenter and the generator.
The cross entropy loss between $S (I_h;\theta_S)$ and $M_h$, where $M_h$ are all zero, is added to $\mathcal{L}_S$.

\begin{figure}[]
    \centering
    \includegraphics[scale = 0.4]{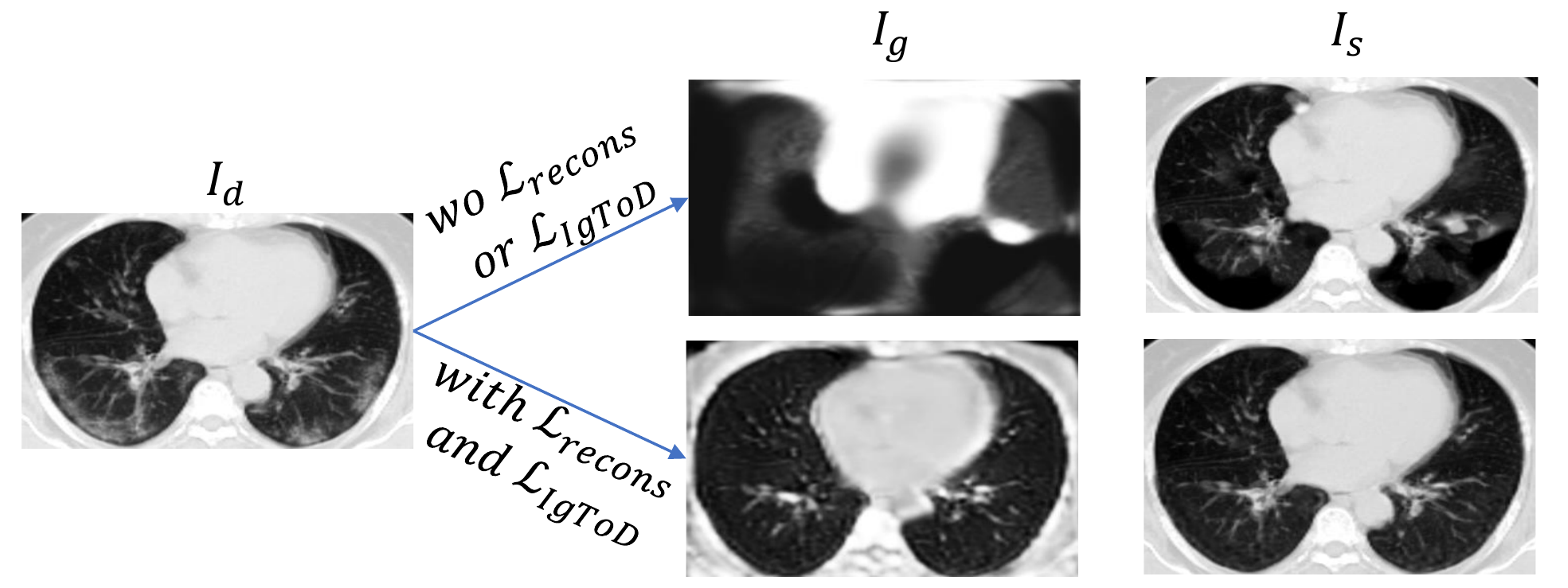}
    \caption{Comparison of the generated volume $I_g$ and the synthetic volume $I_s$ before and after adding  $\mathcal{L}_{recons}$ and $\mathcal{L}_{IgToD}$ of a random COVID-19 CT volume $I_d$ during the training.}
    \label{fig:comparison}
\end{figure}

Second, no supervision constrains the parts of the generated volume where the segmentation values are close to zero
because they are not used to form the synthetic volume, and the quality of the generated volume is uncontrollable due to the lack of the supervision signal.
It becomes the bottleneck of improving the final performance of the segmenter. 
A reconstruction loss $\mathcal{L}_{recons}$ constrains the output of the generator:
$\mathcal{L}_{recons} = \text{MSE} (G (I_h;\theta_G),I_h)$,
where MSE(·,·) is the mean square error function, alleviates the problem.
We also introduce an additional loss 
    $\mathcal{L}_{IgToD} = \mathbb{E}_{\bf{I_d}}[\text{log} (D (G (I_d;\theta_G);\theta_D)]$
to $\mathcal{L}_{GAN}$ for further improvement by feeding the generated volume of $I_d$ into the D.
Fig.~\ref{fig:comparison} shows the comparison of the generated volumes before and after adding $\mathcal{L}_{recons}$ and $\mathcal{L}_{IgToD}$. We can see that the quality of the generated volume and the synthetic volume are
significantly improved, so as to the performance of the segmenter, which will be detailed in section~\ref{sec:ablation}. 
When the input of the S are healthy volumes, the forward propagation process of synthesis and discrimination 
is not needed, and $\mathcal{L}_{GAN}$ will not be calculated.

\begin{figure}[]
    \centering
    \includegraphics[scale = 0.33]{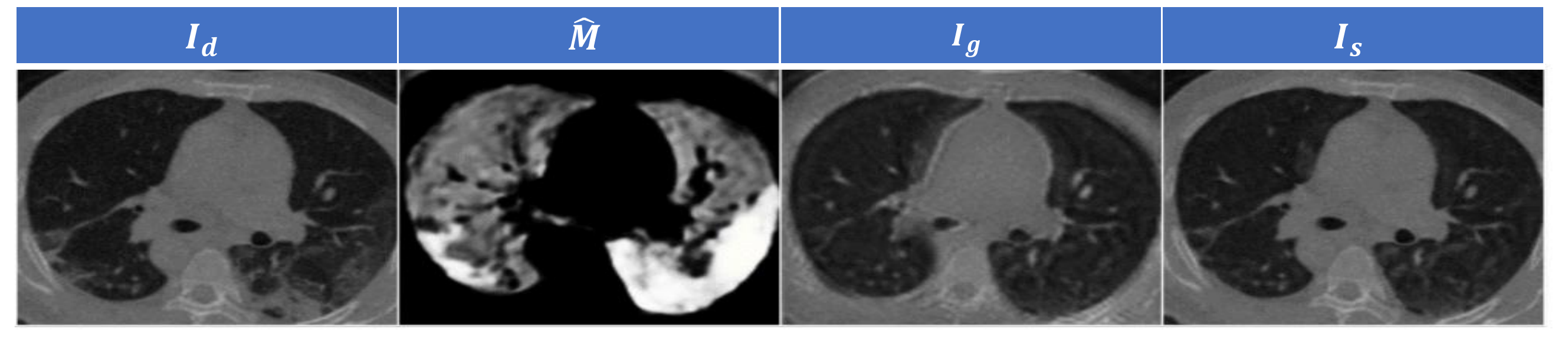}
    \caption{An example of performance collapse during training. $I_d$ is a COVID-19 volume without voxel-level annotation.
    Although the synthetic volume $I_s$ is quite similar to a real healthy volume, 
    the segmentation mask $\hat{M}$ contains a large number of non-lesion regions, indicating a performance collapse.
    }
    \label{fig:collapse}
\end{figure}

As the training proceeds and the synthetic volume gets closer to the real healthy volume, the lesion signal that can be captured by the D becomes weaker and weaker. The D will tend to learn the noise
signal between the data $I_s$ and $I_h$ rather than the pathological signals, which leads to the performance collapse of the GASNet. Fig.~\ref{fig:collapse} gives an example 
where performance collapse happens during training. The segmenter not only segmented the lesion area but also the healthy area, modifying both the pathological signals and 
the noise signals of the synthetic volume to confuse the discriminator. This leads to an extremely low segmentation performance.

\begin{figure}[]
    \centering
    \includegraphics[scale = 0.25]{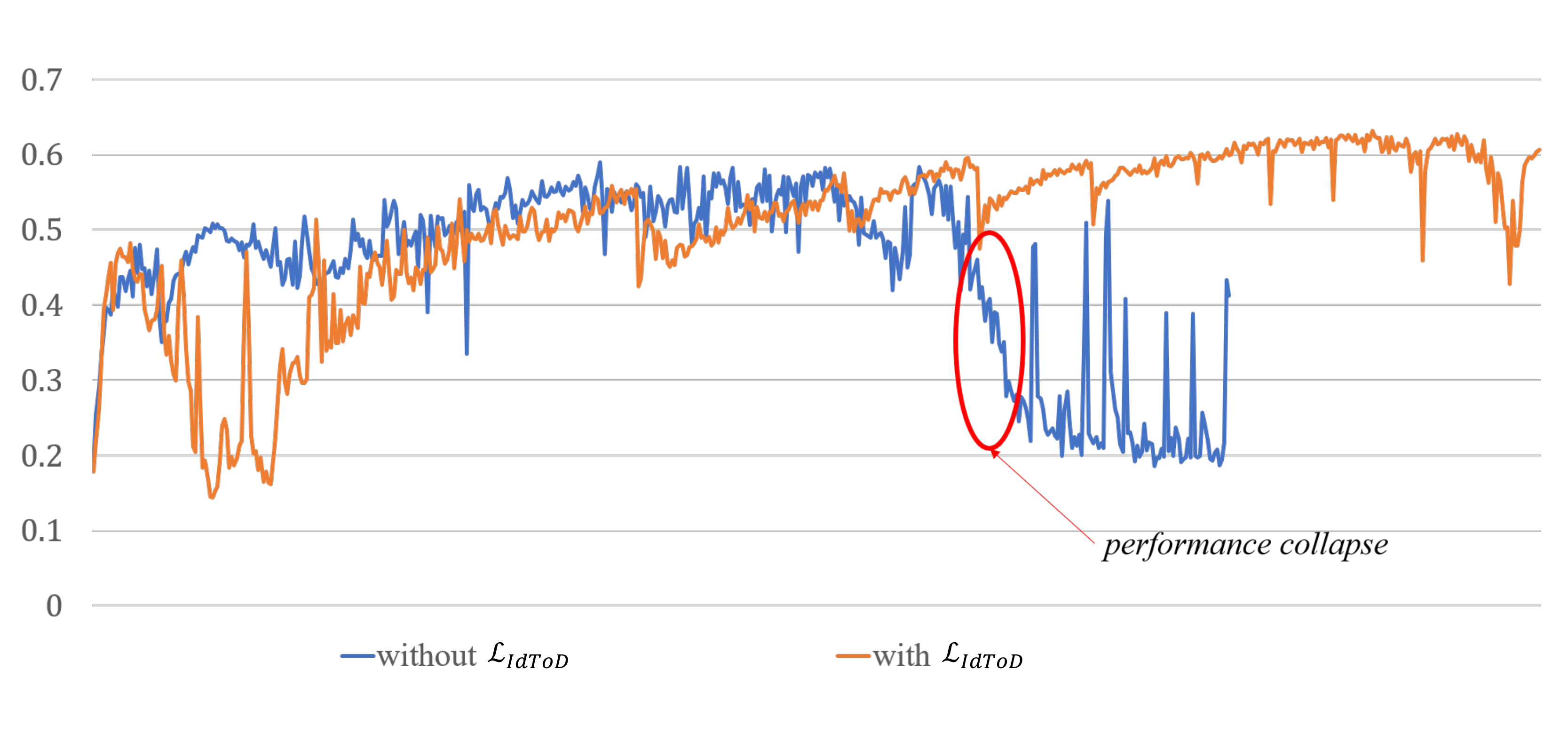}
    \caption{Performance change curve during training w/o $\mathcal{L}_{IdToD}$. The ordinate represents the Dice score on the validation dataset. 
    Without $\mathcal{L}_{IdToD}$, performance collapse begins at the location of the red circle.
    }
    \label{fig:curve}
\end{figure}

Inspired by the idea of dropout in the field of weakly supervised localization~\cite{Choe}\cite{Choe2020a}\cite{Liu2020a},
where a dropout layer randomly determines whether to block the distinguishing features coming into the next layer of the classification network,
we also feed the original diseased volume $I_d$ to the D in order to maintain the sensitivity and discriminability of the discriminator to the lesion signal during the training, 
meaning the dropout ratio is fixed at 0.5. A constraint loss 
    $\mathcal{L}_{IdToD} = \mathbb{E}_{\bf{I_d}}[\text{log} (D (I_d;\theta_D)]$
 is added to $\mathcal{L}_{GAN}$, 
hoping that the D can always distinguish between volumes of the patients and the healthy people.
Fig.~\ref{fig:curve} compares the training curves with and without the auxiliary constraint from $\mathcal{L}_{IdToD}$ and shows that $\mathcal{L}_{IdToD}$ alleviates 
the performance collapse of GASNet markedly.


Finally, as the data $I_d$ has no voxel-level annotations, the final S may segment any lesion areas for some mild infected CT volumes.
Inspired by MIL~\cite{zhou2009multi}, we add a MIL loss to $\mathcal{L}_S$:
$\mathcal{L}_{MIL} = -\text{log} (\text{max} (S (I_d;\theta_S)))$, meaning at least one voxel of an diseased volume should be predicted as positive.

To sum up, we extend loss functions $\mathcal{L}_{GAN}$ and $\mathcal{L}_S$ by adding four new losses as auxiliary constraints. 
The final objective function is defined as follow:
\begin{align*}
    &\quad\mathop{\text{min}}_{\theta_{G},\theta_S} \mathop{\text{max}}_{\theta_D}  \mathcal{L}_{GAN} + \lambda_S \mathcal{L}_S \nonumber \\
&= \mathbb{E}_{\bf{I_h}}[\text{log} (1-D (I_h;\theta_D))]+\mathbb{E}_{\bf{I_d}}[\text{log} (D (I_s;\theta_D))] \nonumber \\
 &+ \underbrace{\mathbb{E}_{\bf{I_d}}[\text{log} (D (I_d;\theta_D)]}_{\mathcal{L}_{IdToD}} 
+ \underbrace{\mathbb{E}_{\bf{I_d}}[\text{log} (D (G (I_d;\theta_G);\theta_D)]}_{\mathcal{L}_{IgToD}} \nonumber \\ 
& +\underbrace{\text{MSE} (G (I_h;\theta_G),I_h)}_{\mathcal{L}_{recons}} + 
\lambda_S \Big( \underbrace{-\text{log} (\text{max} (S (I_d;\theta_S)))}_{\mathcal{L}_{MIL}} \nonumber \\
&+ \text{CSL} (S (I_l;\theta_S),M_l)  +\text{CSL} (S (I_h;\theta_S),M_h) \Big) \nonumber\\
 \nonumber
\end{align*}

\begin{algorithm}[h]
    \caption{Procedure of training GASNet.}
    \label{alg:algorithm1}
    \KwIn{ 
            $\text{diseased volume dataset} \gets Ds_d$\\
            $\qquad\text{healthy volume dataset} \gets Ds_h$\\
            $\qquad\text{labeled volume dataset} \gets Ds_l$\\
            hyperparameters:
            $Repeat_G,Repeat_D,Iter_{ps},Max_{iter}$} 
    \KwOut{GASNet}  
    \BlankLine
    Initialize  the parameters of the segmenter (S), the discriminator (D), and the generator (G);\\
    Initialize iter as 1 \\
    \While{\textnormal{iter $\leq Iter_{ps}$}}{
        \ForEach{$k\in [1,Repeat_G]$}{
            freeze the D, unfreeze the S, the G;\\
            sample $I_d$ from $Ds_d$, $I_h$ from $Ds_h$ \\
            calculate $\mathcal{L}_{GAN}$;\\
            sample $I_l$ and $M_l$ from $Ds_l$, $I_h$ from $Ds_h$ \\
            calculate $\mathcal{L}_S$;\\
            optimize the S, the G;
        }
        \ForEach{$k\in [1,Repeat_D]$}{
            freeze the S, the G, unfreeze the D; \\
            sample $I_d$ from $Ds_d$, $I_h$ from $Ds_h$\\
            calculate $\mathcal{L}_{GAN}$;\\
            optimize the D;
        }
        iter = iter+1
    }
    \While{\textnormal{iter $\leq Max_{iter}$}}{
        repeat 4 to 18 but insert \\
        synthesize $I_{ps}$ and $M_{ps}$ from $I_d$, $I_h$ and $S(I_d)$ \\
        betweent 7 and 8
    }
\end{algorithm}
\subsection{Synthesize COVID-19 positive CT volumes with voxel-level pseudo-label}
\begin{figure}[]
    \centering
    \includegraphics[scale = 0.3]{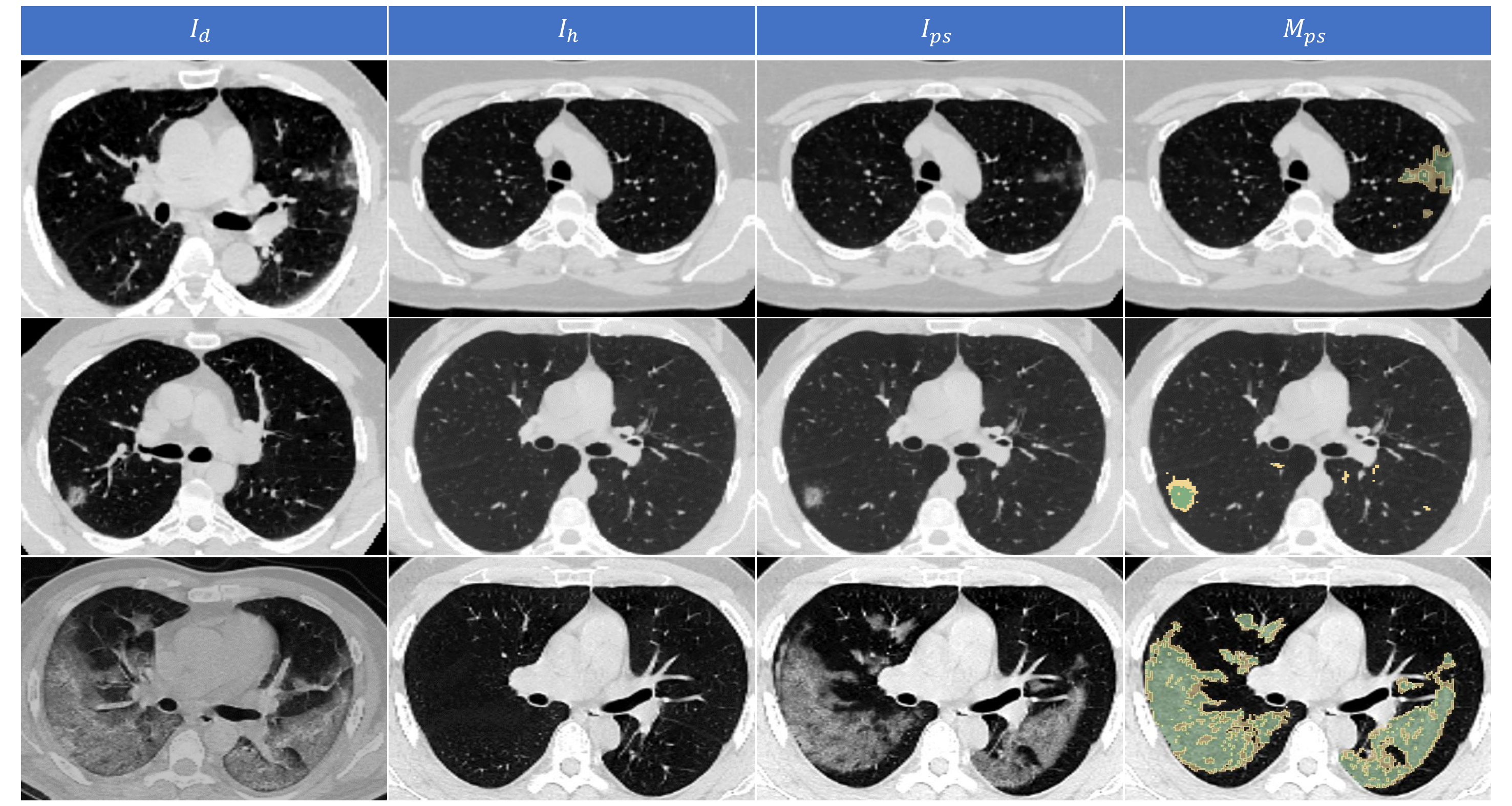}
    \caption{Three examples of the generated COVID-19 positive CT volumes and their pseudo labels.
    From the first column to the fourth column are real COVID-19 CT volumes $I_d$, real healthy CT volumes $I_h$, synthesized COVID-19 CT volumes $I_{ps}$, and pseudo labels $M_{ps}$.}
    \label{fig:sysimage}
\end{figure}

With the losses detailed in subsection~\ref{sec:prior constrain}, GASNet can be trained stably and achieve good performance. 
We can further improve the segmentation performance by synthesizing COVID-19 positive CT volumes with voxel-level pseudo-label during the training process.
Given an unlabeled COVID-19 data $I_d$ and its predicted lesion segmentation mask $\hat{M}=S (I_d;\theta_S)$, 
a healthy data $I_h$ and its predicted lung mask $M_{lung}$ where $M_{lung}$ can be obtained by existing automatic algorithms~\cite{hofmanninger2020automatic}, 
we can synthesize a COVID-19 positive volume $I_{ps}$ and its corresponding pseudo-label $M_{ps}$ as follows:
\begin{equation}
    I_{ps} = I_h\times (1-\hat{M}\times M_{lung})+I_d \times \hat{M}\times M_{lung} \nonumber
\end{equation} 
$$
    M_{ps}^{ij}=\left\{
    \begin{aligned}
    1&,& \text{if } \hat{M}^{ij}\times M_{lung}^{ij} \ge 0.5+\xi \\
    0&,& \text{if } \hat{M}^{ij}\times M_{lung}^{ij} \le 0.5-\xi \\
    2&,&\text{otherwise}
    \end{aligned}
    \right.
$$
where label 2 means voxels whose labels are not considered. We set $\xi$ as $0.3$ in our experiment. 
\begin{figure}[]
    \centering
    \includegraphics[scale = 0.4]{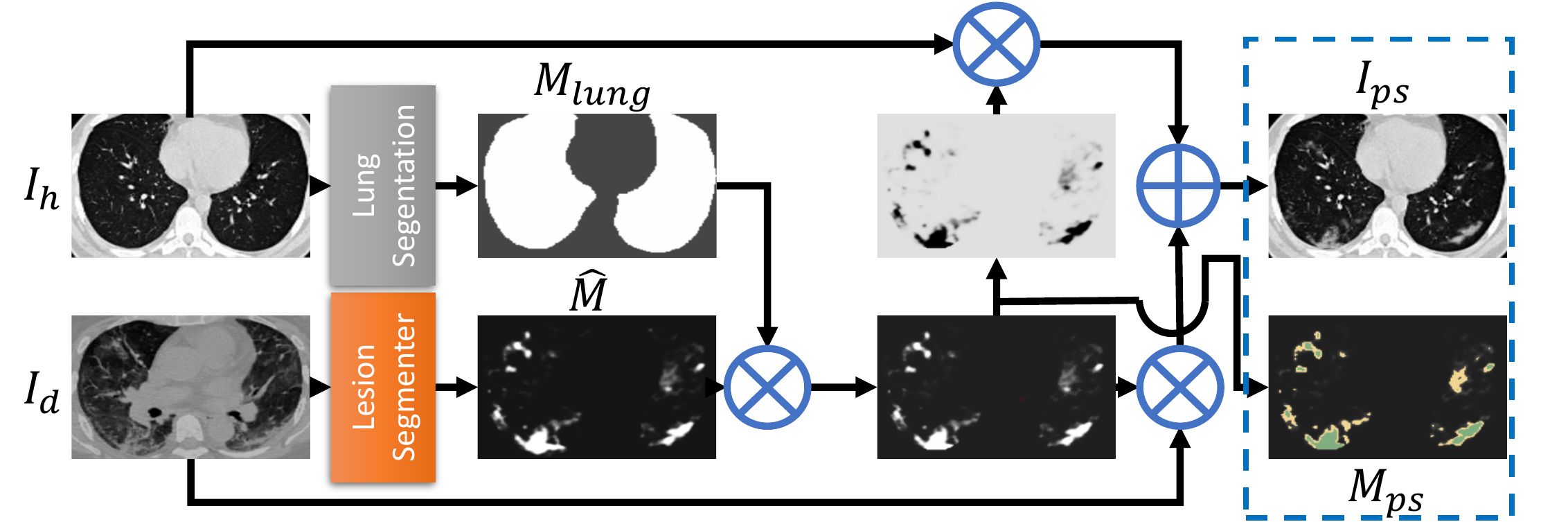}
    \caption{Pipeline of synthesizing COVID-19 positive CT volumes $I_{ps}$ and their pseudo labels $M_{ps}$.
    Real healthy CT samples $I_h$ together with their lung segmentation masks $M_{lung}$, and COVID-19 CT samples $I_d$ together with their predicted lesion segmentation masks $\hat{M}$ are needed.
    The green area of $M_{ps}$ represents the lesion (1), 
    the yellow area is the part ignored when calculating loss (2), and the other areas represent the background (0). 
    The last column in Fig.~\ref{fig:sysimage} follows the same rule.}
    \label{fig:syspipeline}
\end{figure}
Different from the synthesis method in~\cite{Yao2020}, in which the distribution and shpae of lesions added to the healthy volumes are artificailly defined, 
the lesion area of our synthetic data is dynamically extracted from the real COVID-19 positive volumes.
Different from~\cite{Zhang2020a}, in which the synthetic volume is generated through complex cascade generative networks given a label map of lesion and lung, 
our synthetic data is formed by simple linear weighted fusion of real infectious areas and real health volumes.
Fig.~\ref{fig:sysimage} gives three examples of $I_{ps}$ and corresponding $M_{ps}$. 
The synthetic COVID-19 volumes look very natural and diverse. 
Relying on the generated paired data $I_{ps}$ and $M_{ps}$, 
we can alleviate the problem of insufficient voxel-level labeled samples by adding corresponding voxel-level cross entropy loss 
    $\mathcal{L}_{ps} = \text{CSL} (S (I_{ps}; \theta s), M_{ps})$
to $\mathcal{L}_S$ during the training. $\mathcal{L}_{ps}$ boosts the segmentation performance of GASNet by 5.5\% in our experiments, as shown in section~\ref{sec:ablation}.

\subsection{Implementations}
\paragraph{Data pre-processing}
The 3D volume sample of each CT is cropped along the lung mask.
The cropped volume is then resized into $40\times160\times160$. Following the advice from~\cite{COVID-19-SegBenchmark}, we clipped the value of CT volumes into [-1250,250].
As the tanh operation is used as the output of the generated volume, and the value of the output after a tanh operation ranges from - 1 to 1, 
we normalized the input volume into the same range.
The automatic lung segmentation algorithm is based on an open-source pre-trained U-Net model~\cite{hofmanninger2020automatic}.
This lung segmentation algorithm may not be perfect in some cases, but we just use it to get an approximate bounding box around the lung area.
Lung segmentation of COVID-19 volumes are never used after pre-processing.
\paragraph{The structure of GASNet}
Without loss of generality, we adopt the standard U-Net structure as the segmenter of GASNet.
Regarding the memory usage of the 3D volume,
the number of the basic channel is reduced to 16 from 64 in the original paper.
The generator and discriminator follows the structure of CycleGAN~\cite{zhu2017unpaired}.
Following the advice in~\cite{miyato2018spectral},
we add the spectral normalization operation to the discriminator.
\paragraph{Training strategy and hyperparameters}
Four datasets are required in order to train GASNet and get the best model, as shown in Algorithm~\ref{alg:algorithm1}. 
Since the target between training the S and the G and training the D is adversarial, we iteratively update their parameters
using the corresponding loss in each step.
Both $\mathcal{L}_{GAN}$ and $\mathcal{L}_S$ have contributions to the optimization of the segmenter,
so there exists a hyperparameter $\lambda_s$ to balance the two losses:
$\mathcal{L}_{GAN}+\lambda_s\mathcal{L}_S$.
As G and D are trained alternately,
a hyperparameter $\theta_i$ controls the ratio of the number of times G and D are trained in each alternation. Validation is carried out 
every $Val_{iter}$ and only the parameters with the best performance on the validation dataset will be saved. 

\section{Experiments}
\label{sec:experiment}
\subsection{Dataset}
We test the performance of our method on three public COVID-19 CT segmentation datasets~\cite{MedSeg}\cite{morozov2020mosmeddata}\cite{COVID-19-SegBenchmark}.
Another public dataset with only slice-level annotations~\cite{SIRM} used in~\cite{fan2020inf}\cite{qiu2020miniseg} is not suitable for GASNet for two reasons: 
(1) GASNet takes 3D CT volumes, rather than 2D slices as input; (2) 
annotations on slice-level, indicating whether a slice contains lesion area, are not directly available from diagnosis results.   

\textbf{Dataset-A}~\cite{COVID-19-SegBenchmark} consists of 20 CT volumes. 
Lungs and areas of infection were labeled by two radiologists and verified by an experienced radiologist. CT values of 10 volumes have been transformed to the range of [0,255]. 
Considering original CT values are unavailable, some work~\cite{Yao2020} 
did not test the performances on these volumes. We divide the dataset
into $\text{subset}_1$ (original CTs) and $\text{subset}_2$ (10 transformed CTs),
like~\cite{Yao2020}\cite{Zhang2020a}, and report the separate and overall performances.

\textbf{Dataset-B}~\cite{MedSeg} consists of 9 COVID-19 CT volumes with voxel-level annotations by a radiologist.

\textbf{Dataset-C} and \textbf{Dataset-D (Volume-level annotation)} are from MosMed~\cite{morozov2020mosmeddata}, 
which consists of 856 CT volumes with COVID-19 related findings as well as 254 CT volumes without such findings.
50 COVID-19 cases have voxel-level annotations of lesions by experts, which forms Dataset-C.
The rest of the data, consisting of 254 healthy volumes and 806 COVID-19 volumes excluding 50 voxel-level labeled samples, forms Dataset-D. 
The diagnosis results of the CT volumes can be used as volume-level labels directly.

\textbf{Dataset-E (Volume-level annotation)} is
a large dataset with volume-level annotation we collected, in which
1,678 COVID-19 CT volumes come from the Wuhan Union Hospital,
whose patients have been diagnosed as COVID-19 positive by nucleic acid testing,
and 1,031 healthy CT volumes come from the routine physical examination.

\subsection{Experimental settings}
For training, all volume-level labeled data in Dataset-E is used to optimize GASNet.
As for voxel-level labeled data, one volume randomly selected from Dataset-A is used for training and all of the rest, including 19 cases of Dataset-A and all 
volumes from Dataset-B and Dataset-C, are used for the test. 
Since Dataset-D comes from the same source as Dataset-C, we finetune GASNet using the volume-level Dataset-D when testing the performance on Dataset-C. 
The finetuned model is marked as $GASNet_{finetune}$.

As for the hyperparameters, $\lambda_s$ is set to 100; $\theta_i$ is set to 5, meaning GASNet optimizes D 5 times each time it optimizes S and G.
GASNet is trained jointly from scratch (without pre-training),
with a batch size of 4, learning ratio of 1e-5 for the D and the S and 1e-4 for the S. 
$\mathcal{L}_{ps}$
is not calculated in the first 7,000 iterations as we found the predicted mask for $I_d$ is prone to errors at first.   
Simple data augmentation techniques, including random cropping, Gaussian noise, and rotation 
lead to slight improvement on the test dataset.
It takes about 24 hours ($\sim$14,000 iterations) for training using a Titan RTX GPU with a 24G memory. 
During the test, voxels greater than 0.5 in the probabilistic segmentation mask $\hat{M}$ are predicted to be lesion (1), and those smaller than 0.5 are predicted to be healthy (0).


\subsection{Results}
We adopt typical metrics in COVID-19 lung infection quantification~\cite{shi2020large}\cite{shan2020lung}, i.e. the Dice Score,
Sensitivity, and Specificity for evaluation.

\textbf{Dice Score} measures the overlap
between the prediction and the ground truth: $Dice = \frac{2\times TP}{2\times TP+FP+FN}$, where TP, FP, and FN are the number of true
positive, false positive, and false negative voxels of one CT volume. 

\textbf{Sensitivity} measures the fraction of real positive samples that are predicted correctly:
$Sensitivity = \frac{TP}{TP+FN}$. 

\textbf{Specificity} measures the fraction
of real negative samples that are predicted correctly:
$Specificity = \frac{TN}{FP+TN}$.

Quantitative results of GASNet and other small sample learning work on COVID-19 segmentation testing on three public datasets are shown in Table~\ref{tab:dataset-A} and~\ref{tab:dataset-BC}.
We also reproduce the MIL strategy to represent the mainstream weakly-supervised methods in general medical image segmentation, together with a standard segmentation network, 
which is a U-Net structure in our experiment.
Because different methods used inconsistent division strategies for datasets, the tables also show the number of training samples and testing samples used by each method on each dataset.
In order to understand the difficulty of COVID-19 lesion segmentation, 
two radiologists from Wuhan Union Hospital annotated cases of Dataset-A in voxel level independently and their performances are measured by comparing with the ground truth of Dataset-A. 
The Dice scores of two radiologists are $73.5\%$ and $73.9\%$, while GASNet achieves $70.3\%$.
Comparing with other existing works, only LabelFree~\cite{Yao2020} and CoSinGAN~\cite{Zhang2020a} use less voxel-level labeled samples in training (zero and one slice from one sample) than GASNet, 
but GASNet exceeds their performance with a large margin. 
Other methods including~\cite{Laradji2020}\cite{laradji2020weakly} and 3D nnU-Net~\cite{isensee2018nnu} from the SegBenchmark~\cite{COVID-19-SegBenchmark}
use more training samples and their performance can not match that of GASNet. 

\begin{figure}[]
    \centering
    \includegraphics[scale = 0.3]{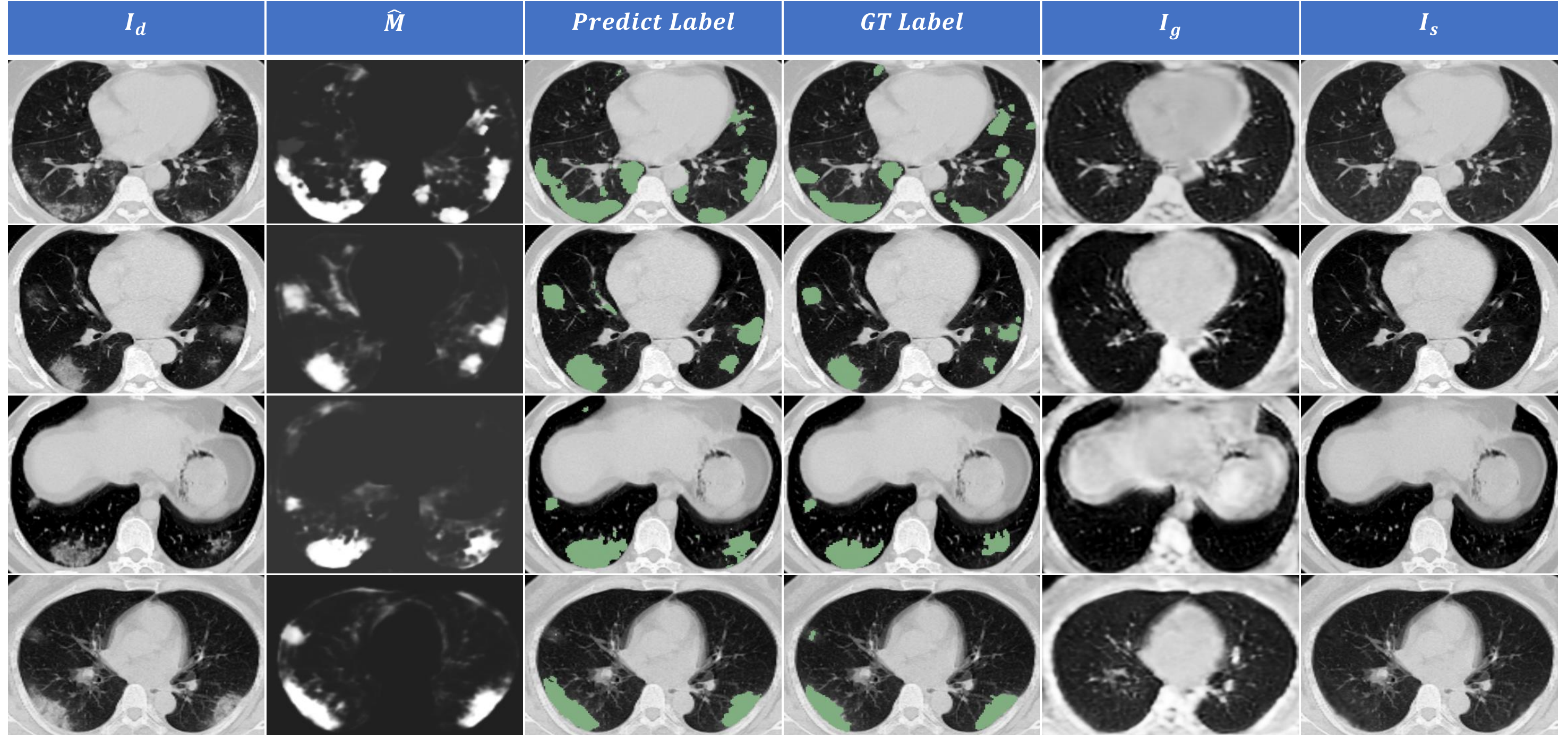}
    \caption{The segmentation and intermediate results of GASNet of four test examples in Dataset-A. 
    From the first column to the last column are COVID-19 CT volumes $I_d$, predicted
     segmentation masks $\hat{M}$, predicted segmentation results, ground truth of COVID-19 CT volumes, generated CT volumes $I_g$, and synthesized CT volumes $I_s$.
    }
    \label{fig_vis}
\end{figure}

Qualitatively, visualization of the output of GASNet on four samples from the test datasets is shown in Fig.~\ref{fig_vis}. 
The generated volume looks like a blurry version of the original input, except for the predicted 
lesion areas. The appearance in the predicted lesion areas changes a lot, making the generated volume look closer to a real healthy volume. Therefore, GASNet replaces the lesion areas of the original
diseased volume with the corresponding parts of the generated volume, which makes the synthetic volume look quite similar to real healthy volumes. These examples show that GASNet does optimize its parameters to 
reach the goal of restoring original healthy CT volumes as 
we expect.

\begin{figure}[]
    \centering
    \includegraphics[scale = 0.3]{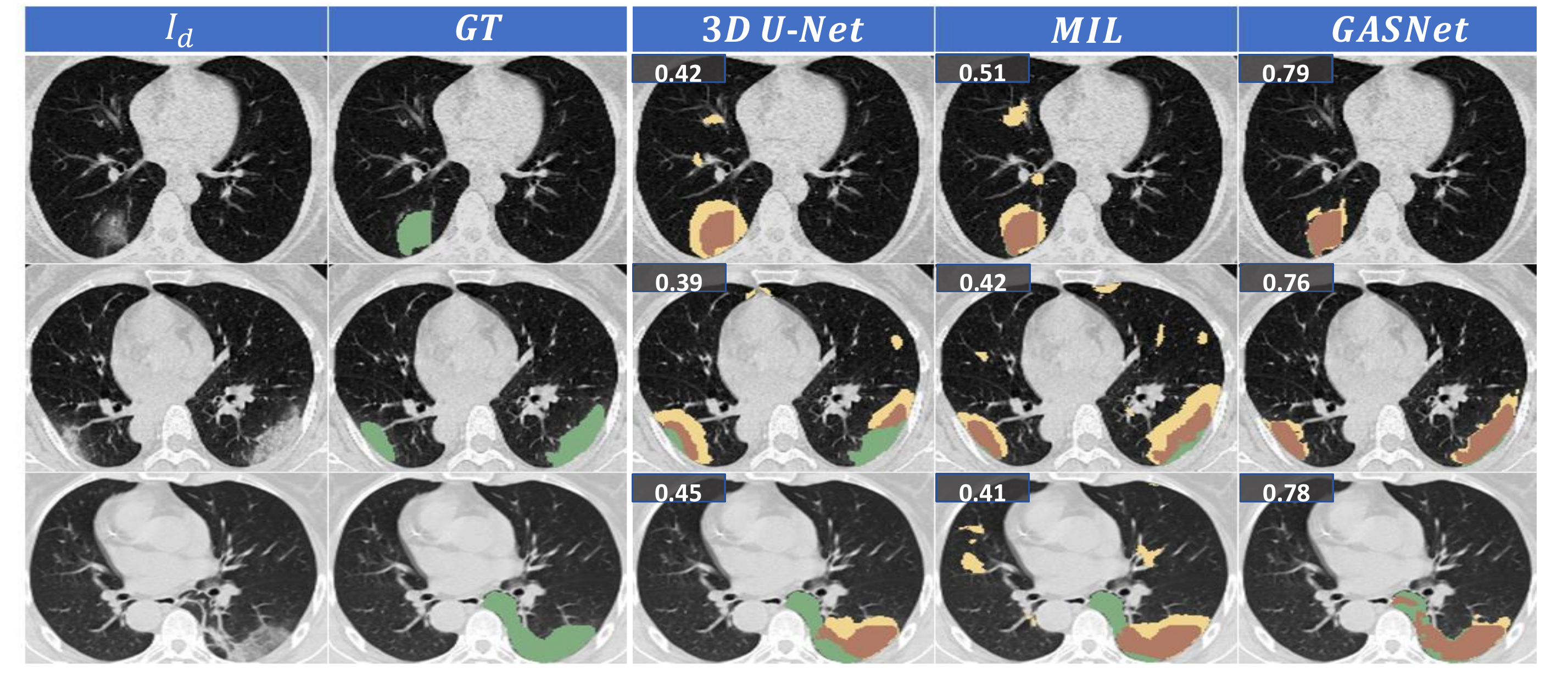}
    \caption{Segmentation results of three test samples in Dataset-A by three different segmentation algorithms. 
    All algorithms used one voxel-level labeled sample for training.
    $I_d$ represents COVID-19 CT volumes ; $GT$ represents ground truth.
    The last three columns represent three different segmentation results by 3D U-Net, MIL, and GASNet. 
    The green, yellow, and brown areas in the last three columns represent false negative, false positive, and true positive respectively.
    Numbers in the upper left corner represent Dice scores of current segmentation results.
    }
    \label{fig_vis2}
\end{figure}

\begin{figure}[]
    \centering
    \includegraphics[scale = 0.3]{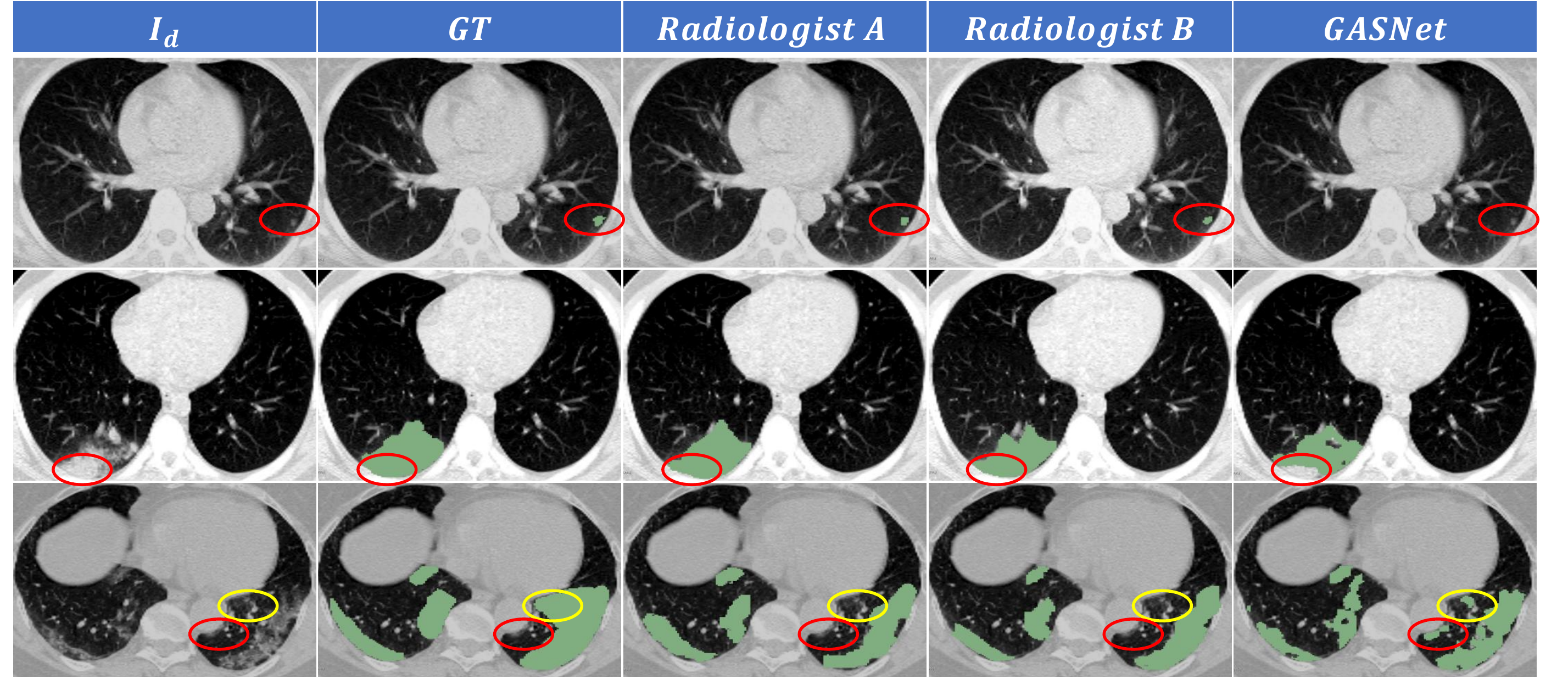}
    \caption{Three unsatisfactory cases. 
    Red circles circle the unsatisfying parts of predicted segmentation,
    where GASNet predicts inconsistent segmentation with annotations of the two radiologists and the ground truth.
    Yellow circles circle the controversial parts. 
    In these regions, annotations of radiologists and the ground truth are inconsistent, 
    while the segmentation of GASNet is closer to those of radiologists.
    }
    \label{fig_vis4}
\end{figure}

The segmentation results of three samples using different methods are shown in Fig.~\ref{fig_vis2}.
Compared with the standard 3D U-Net baseline and Multi-Instance Learning method, 
GASNet holds obvious advantages in eliminating both false positive and false negative. 
Fig.~\ref{fig_vis4} shows three cases where GASNet has relatively poor performance.
The first CT volume contains a small lesion, while GASNet missed it. 
In the second case, the lesion segmentation of GASNet is partially missing near the edge of lung. 
In the third case, the lesion area is so complex that 
annotations of the two radiologists and the ground truth are inconsistent, 
while the segmentation of GASNet is closer to those of radiologists.

\begin{table*}[htb]
    \caption{Quantitative results of COVID-19 segmentation on the dataset-A.}
    \centering
    \resizebox{1.98\columnwidth}{!}{
    \begin{tabular}{lccccccccccc}
        \hline
        \multirow{2}{*}{Method}  & \multirow{2}{*}{Training cases} & \multirow{2}{*}{Testing cases} & 
        \multicolumn{3}{c}{$\text{subset}_1$} & \multicolumn{3}{c}{$\text{subset}_2$} & \multicolumn{3}{c}{whole}  \\
        \cline{4-12}
        &&& $Dice$ $\%$ & $Sensitivity$ $\%$ & $Specificity$ $\%$ & $Dice$ $\%$ & $Sensitivity$ $\%$ & $Specificity$ $\%$ & $Dice$ $\%$ & $Sensitivity$ $\%$ & $Specificity$ $\%$ \\
        \hline
        ActiveLearning\cite{Laradji2020} & 16 (point-label) & 4        & - & - & - & - & - & - & 44 & - & - \\
        3D nnU-Net\cite{isensee2018nnu} & 4 & 16         & - & - & - & - & - & - & $67.3\pm22.3$ & - & - \\
        CoSinGAN\cite{Zhang2020a} & 1  & 19               & $57.8$ & - & - & 48.4 & - & - & 54.8 & - & -  \\
        LabelFree\cite{Yao2020} & 0 & 20                 & $68.7\pm15.8$ & $62.1\pm22.8$ & - & - & - & - & - & - & - \\
        MIL\cite{Xu} & 1 & 19                 & $51.6\pm14.1$ & $84.3\pm10.1$ & $97.3\pm1.3$ & $44.5\pm22.7$ & $53.3\pm27.3$ & $98.0\pm0.9$ & $48.1\pm19.2$ & $68.8\pm25.8$ & $97.6\pm1.2$ \\
        MIL\cite{Xu} & 4 & 16                 & $61.8\pm8.1$ & $76.4\pm16.8$ & $98.9\pm0.4$ & $49.0\pm19.7$ & $\bf{60.2\pm22.8}$ & $99.1\pm0.8$ & $55.4\pm16.4$ & $68.3\pm21.6$ & $99.0\pm0.7$ \\
        \hline
        $\text{GASNet}$ & 1 & 19                                 & $\bf{76.7\pm6.1}$ & $\bf{84.6\pm7.2}$ & $\bf{99.2\pm0.7}s$ & $\bf{63.2\pm19.4}$ & $58.4\pm25.2$ & $\bf{99.6\pm0.4}$ & $\bf{70.3\pm17.1}$ & $\bf{70.0\pm21.8}$ & $\bf{99.4\pm0.6}$ \\
        radiologist A & - & 20                                 & - & - & - & - & - & - & $73.9\pm18.0$ & $66.2\pm20.1$ & $99.8\pm0.2$ \\
        radiologist B & - & 20                                 & - & - & - & - & - & - & $73.5\pm21.1$ & $71.7\pm21.2$ & $99.6\pm0.4$ \\
        \hline
    \end{tabular}
    }
    \label{tab:dataset-A}
\end{table*}
\begin{table*}[htb]
    \caption{Quantitative results of COVID-19 segmentation on the dataset-B and dataset-C.}
    \centering
    \resizebox{1.98\columnwidth}{!}{
    \begin{tabular}{lcccccccc}
        \hline
        \multirow{2}{*}{Method}  & \multirow{2}{*}{Training cases} & \multirow{2}{*}{Testing cases} & 
        \multicolumn{3}{c}{Dataset-B} & \multicolumn{3}{c}{Dataset-C}  \\
        \cline{4-9}
        &&& $Dice$ $\%$ & $Sensitivity$ $\%$ & $Specificity$ $\%$ & $Dice$ $\%$ & $Sensitivity$ $\%$ & $Specificity$ $\%$ \\
        \hline
        ActiveLearning\cite{Laradji2020} & 6 (point-label) & 3      & 52.4 & - & -            & - & - & - \\
        3D nnU-Net\cite{isensee2018nnu}  & 4 (from Dataset-A) & 50      & - & - & -               & $58.8\pm20.6$ & - & -  \\
        LabelFree\cite{Yao2020}          & 0 & 8      & $59.4\pm17.4$ & $61.8\pm18.4$ & -          & - & - & - \\
        MIL\cite{Xu}          & 4 (from Dataset-A) & 9+50      & $43.7\pm19.5$ & $69.0\pm24.6$ & $98.8\pm1.0$          & $34.9\pm20.5$ & $52.3\pm25.6$ & $99.3\pm0.4$ \\
        \hline
        $\text{GASNet}$ & 1 (from Dataset-A) & 9+50    & $\bf{60.2\pm23.4}$ & $\bf{66.8\pm28.9}$ & ${99.3\pm0.5}$          & ${54.2\pm22.4}$ & $55.6\pm28.3$ & $99.6\pm0.2$ \\
        $\text{GASNet}_{finetune}$ & 1 (from Dataset-A) & 9+50    & $59.7\pm18.5$ & $66.5\pm26.3$ & $\bf{99.3\pm0.2}$          & $\bf{58.9\pm24.4}$ & $\bf{60.4\pm27.5}$ & $\bf{99.8\pm0.2}$ \\
        \hline    
    \end{tabular}
    }
    \label{tab:dataset-BC}
\end{table*}

\section{Ablation Study}
\label{sec:ablation}

\subsection{Number of the voxel-level labeled samples}
To understand the impact of the number of training samples with voxel-level annotations, we use 1 sample, 4 samples, 20 samples and 45 samples respectively from  Dataset-C as voxel-level labeled samples, 
use  Dataset-D as volume-level labeled samples to train GASNet, and test the performance their performance on Dataset-A.  
We also train the baseline, i. e., using only a corresponding number of voxel-level labeled samples to train the standard U-Net network.

\begin{figure}[]
    \centering
    \includegraphics[scale = 0.65]{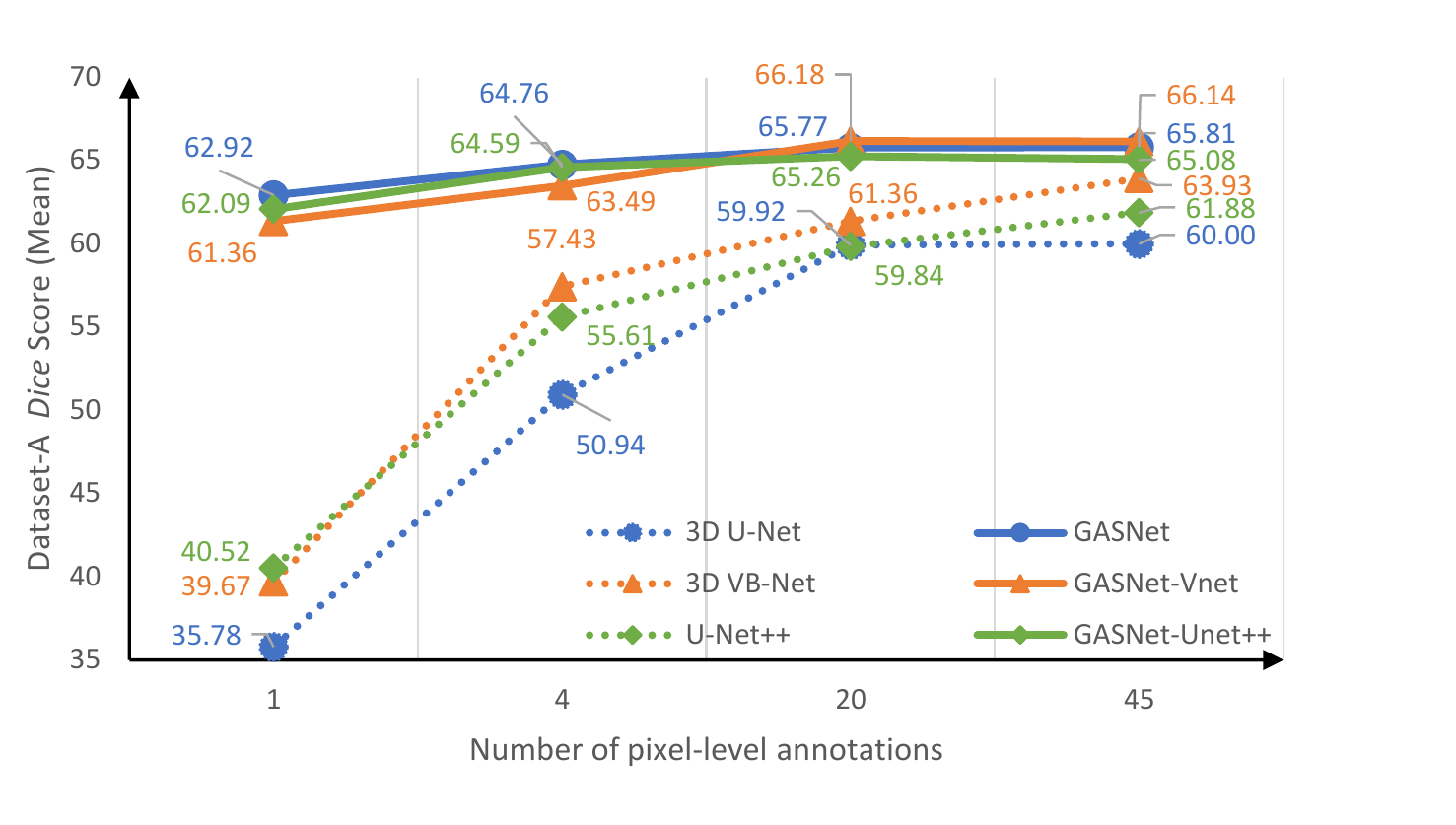}
    \caption{Segmentation performances of six algorithms (3D U-Net, 3D VB-Net, U-Net++, and corresponding GASNet versions) 
    on Dataset-A using different numbers of training samples with voxel-level annotations. 
    }
    \label{fig:performance}
\end{figure}

Besides of 3D U-Net, we also
adopt 3D VB-Net~\cite{milletari2016v} and U-Net++~\cite{zhou2018unet++} as the segmenter of GASNet and test the performance following the same experimental scenario.
The Dice scores on the test dataset of all the experiments are shown in Fig.~\ref{fig:performance}. 
Note that 3D VB-Net is also the network used in~\cite{shan2020lung}.
The performance of GASNet, no matter which segmentation model is used as the segmenter of GASNet, is always better than that of the corresponding baseline, which demonstrates the robustness of the framework.

\subsection{Improvement for segmentation performance by different auxiliary loss functions and $\mathcal{L}_{ps}$}
As demonstrated in subsection~\ref{sec:prior constrain}, auxiliary constraints added as loss functions benefit the training of GASNet and the final performance. We quantitatively analyzed the ability of different constraints 
to improve the final segmentation performance by gradually adding the constraint losses to the framework. The quantitative results are shown in Table~\ref{ablation-augmentation}. Each auxiliary constraint benefits the performance,
with $\mathcal{L}_{IgToD}$ and $\mathcal{L}_{IdToD}$ benefiting the most. As shown in Fig.~\ref{fig:comparison} and Fig.~\ref{fig:curve}, $\mathcal{L}_{IgToD}$ improves the quality of the generated volume and $\mathcal{L}_{IdToD}$ alleviates the 
performance collapse of GASNet. Compared with the original GASNet without any auxiliary constraints, the Dice score has cumulatively risen more than 10 percent, proving the great impact of auxiliary constraints on network training.
$\mathcal{L}_{ps}$ further improves the segmentation performance of GASNet from 64.75\% to 70.3\%, by adding reliable supervision signal in voxel-level to the segmenter of GASNet. 
\begin{table}
    \caption{Improvement of segmentation performance by different auxiliary loss functions.}
    \label{ablation-augmentation}
    \centering
    \scriptsize
    \begin{tabular}{cccccc}
      \hline
      $\mathcal{L}_{recons}$ & $\mathcal{L}_{IgToD}$ & $\mathcal{L}_{IdToD}$& $\mathcal{L}_{MIL}$  & $\mathcal{L}_{ps}$  & $Dice$ $ (\%)$\\
      \hline
      $\times$ &  $\times$  & $\times$ & $\times$ &$\times$ &  54.13\\
      \checkmark & $\times$ &  $\times$ &  $\times$ &$\times$ & 56.47 (+2.34) \\
      \checkmark & \checkmark & $\times$  &  $\times$ &$\times$ & 60.67 ($\bf{+4.20}$) \\
      \checkmark & \checkmark & \checkmark &  $\times$  &$\times$ & 64.05 (+3.38) \\
      \checkmark & \checkmark & \checkmark & \checkmark &$\times$ & { $\bf{64.75}$ (+0.70)}\\
      \checkmark & \checkmark & \checkmark & \checkmark &\checkmark & { $\bf{70.3}$ ($\bf{+5.55}$)}\\
      \hline
    \end{tabular}
  \end{table}

\section{Conclusion}
We propose a weakly-supervised framework for COVID-19 infection segmentation, named GASNet.
Utilizing volume-level annotation information,
GASNet needs only a single voxel-level labeled sample to obtain performance comparable to fully-supervised methods. Several auxiliary constraint losses benefit the training of GASNet, improving
the segmentation performance and the quality of the  synthetic volumes. Extensive experiments demonstrate the robustness of the algorithm.
Given that volume-level labels are directly available as diagnosis results,
GASNet is valuable in medical practice. 

However, more research on explaining and improving the framework is necessary,
including 
embedding state-of-the-art segmentation structure to pull up the performance
 and relaxing some constraints used in this study.
In the future, we will try to extend GASNet to handle multi-class segmentation tasks.
Experiments on segmenting lesions of other diseases will also be carried out to valid the generalization of GASNet.

\bibliographystyle{IEEEtran}
\bibliography{IEEEabrv,tmi,weaksupervision,dataset,weakCOVID,dropoutLayer,radio}
\end{document}